\documentclass[twocolumn]{aastex63}
\usepackage{natbib}
\usepackage{empheq}
\usepackage{epstopdf}
\usepackage{subfigure}
\usepackage{amsmath,amsfonts,amssymb}
\usepackage{multirow}
\usepackage{graphicx}
\begin{document}

\title{The Masses of Isolated Neutron Stars Inferred from the Gravitational Redshift Measurements}
\author{Shao-Peng Tang}
\author{Jin-Liang Jiang}
\affil{Key Laboratory of Dark Matter and Space Astronomy, Purple Mountain Observatory, Chinese Academy of Sciences, Nanjing, 210033, People's Republic of China}
\affil{School of Astronomy and Space Science, University of Science and Technology of China, Hefei, Anhui 230026, People's Republic of China}
\author{Wei-Hong Gao}
\affil{Department of Physics and Institute of Theoretical Physics, Nanjing Normal University, Nanjing 210046, People's Republic of China}
\author{Yi-Zhong Fan}
\author{Da-Ming Wei}
\affil{Key Laboratory of Dark Matter and Space Astronomy, Purple Mountain Observatory, Chinese Academy of Sciences, Nanjing, 210033, People's Republic of China}
\affil{School of Astronomy and Space Science, University of Science and Technology of China, Hefei, Anhui 230026, People's Republic of China}
\email{yzfan@pmo.ac.cn (YZF)}

\begin{abstract}
For some neutron stars (NSs) in the binary systems, the masses have been accurately measured. While for the isolated neutron stars (INSs), no mass measurement has been reported yet. The situation will change soon thanks to the successful performance of the Neutron Star Interior Composition Explorer (NICER), with which the radius and mass of the isolated \objectname{PSR J0030+0451} can be simultaneously measured. For most INSs, no mass measurements are possible for NICER because of observational limitations. Benefiting from recent significant progress made on constraining the equation of state of NSs, in this work we propose a way to estimate the masses of the INSs with the measured gravitational redshifts. We apply our method to \objectname{RX J1856.5-3754}, \objectname{RX J0720.4-3125}, and \objectname{RBS 1223}, three members of ``The Magnificent Seven" (M7), and estimate their masses to be $1.24_{-0.29}^{+0.29}M_{\odot}$, $1.23_{-0.05}^{+0.10}M_{\odot}$, and $1.08_{-0.11}^{+0.20}M_{\odot}$, respectively. These masses are consistent with that of binary NS systems, suggesting no evidence for experiencing significant accretion of these isolated objects.
\end{abstract}

\section{Introduction}
Since the discovery of the first pulsar by \citet{1968Natur.217..709H}, more than 2500 neutron stars (NSs) have been detected in the Galaxy and the masses of a small fraction of these objects, usually in NS binary systems, have been accurately measured with some feasible approaches (see a comprehensive review in \citealt{2016ARA&A..54..401O} and references therein). These systems include (1) timing binary pulsars with Keplerian orbit and/or post-Keplerian parameter measurements; (2) millisecond pulsars in globular clusters, or millisecond pulsar-white dwarf systems with measurements of Shapiro delay or with the white dwarf's mass constrained from spectroscopic observations; (3) NSs with a black widow companions \citep{2018ApJ...859...54L} or NSs in stellar triple systems; (4) NSs with high-mass companions or NSs in low-mass X-ray binaries (LMXBs) with Type-I X-ray bursts powered by thermonuclear burning of accreted material.

For isolated neutron stars (INSs) that consist of about $90\%$ of radio pulsars, the mass measurements are much more challenging and there is no report yet. The Neutron Star Interior Composition Explorer (NICER), which in principle can deliver a measurement of radius of NSs with unprecedented accuracy using a pulse profile modeling method \citep{2012SPIE.8443E..13G, 2016ApJ...832...92O, 2019ApJ...880...74M, 2019AIPC.2127b0008W}, is expected to solve this problem. Recently, using more than 1.9 Ms of data acquired for \objectname{PSR J0030+0451}, the NICER team has successfully measured the mass of a solitary NS for the first time and the results are expected to be announced in the future\footnote{\url{https://heasarc.gsfc.nasa.gov/docs/nicer/science_nuggets/20190711.html}}. However, measurements by NICER are heavily dependent on the accumulated observation time/data, pulse profile model, and the brightness of the source. For most INSs, the masses are still unmeasurable.

In this work, we propose a new approach to infer the masses of INSs using the equations of state (EoSs) constrained with gravitational wave (GW) data, nuclear experiments, and the maximum mass of nonrotating NS. This is because with the constrained EoSs, we can map a series of mass\textendash radius ($M \textendash R$) points that were obtained by solving Tolman\textendash Oppenhimer\textendash Volkoff (TOV) equations to the gravitational redshifts ($z_{\rm g}$) and the masses of NS, then we have a $z_{\rm g} \textendash M$ probability distribution (see Fig.\ref{fig:z_Mcurves} below) which effectively bounds the range of the mass of an NS ($M$) for a given $z_{\rm g}$. And with the detection of binary neutron star (BNS) or NS\textendash black hole (NSBH) systems by the LIGO/Virgo Collaboration, the radii of NSs within a wide mass range can be constrained to $\sim$ $10\%$ at $90\%$ confidence \citep{Jiang2019, 2019PhRvD.100j3009H}, thus our method can achieve a reasonably high accuracy and is hence useful for constructing the mass distribution of INSs.

This work is organized as follows: in Section \ref{sec:methods} we introduce the methods of parameterizing EoS, Bayesian parameter inference of GW data, and the approach of estimating the masses of INSs with $z_{\rm g}$. The results on the bulk properties of NSs derived using the posterior samples of parameterized EoSs and masses of INSs inferred from the EoS sets with gravitational redshift measurements are presented in Section \ref{sec:results}. Section \ref{sec:summary} is our discussion and summary.

\section{\label{sec:methods}Methods}
\subsection{\label{sec:para-eos}Parameterizing EoS}
Parameterized representations of the EoS are widely used tools to connect the properties of NSs with the EoS of ultra dense matter, which can be combined with a Bayesian approach to narrow down possible regions of parameter space using observations of NSs and the GW data. A number of parameterizing methods to effectively represent different kinds of realistic EoS models have been developed \citep{2009PhRvD..79l4032R, 2009PhRvD..80j3003O, 2010PhRvD..82j3011L, 2014ApJ...789..127K, 2016EPJA...52...18S, 2017ApJ...844..156R, 2019EPJA...55..209L, 2019PhRvD.100h3010F}. In this work, we will use three methods to parameterize the EoS, including the pressure-based spectral decomposition \citep{2010PhRvD..82j3011L}, the piecewise polytrope parameterization \citep{2009PhRvD..79l4032R}, and a revised version of piecewise polytrope parameterization (i.e., the so-called four-pressure-based piecewise representation; \citealt{Jiang2019}). The first two methods are implemented in the LALSimulation package \citep{2018PhRvD..98f3004C, lalsuite}.

The first method describes the adiabatic index of an EoS as a function of pressure with four expansion coefficients $\{\gamma_0, \gamma_1, \gamma_2, \gamma_3\}$,
\begin{equation}
\label{eq:Gamma_p}
\Gamma(p) = \exp\left[\sum_{k=0}^{3} \gamma_{\rm k} \left(\log\frac{p}{p_0}\right)^{\rm k}\right],
\end{equation}
where $p_0\approx5.3\times10^{32}~{\rm dyn\,cm^{-2}}$ is the reference pressure \citep{2018PhRvD..98f3004C}. We can then construct the EoS from the functional form of the adiabatic index.

The second method, which has four parameters $\{\log_{10}(p_1), \Gamma_1, \Gamma_2, \Gamma_3\}$, however, directly expresses EoS as a relationship between the pressure $p$ and rest-mass density $\rho$ using three segments of polytrope, where $p_1$ is the pressure at density of $1.85\rho_{\rm sat}$ ($\rho_{\rm sat}=2.7\times10^{14}~{\rm g\,cm^{-3}}$ is the so-called nuclear saturation density), $\Gamma_1$ is adiabatic index anchored at ($1.85\rho_{\rm sat}$,$p_1$), $\Gamma_2$ and $\Gamma_3$ are adiabatic indices in the two segments separated by three densities of $\{1.85, 3.7, 7.4\}\rho_{\rm sat}$. The EoS in each segment can be expressed as
\begin{equation}
p(\rho) = K_{\mathit{i}}\rho^{\Gamma_{\mathit{i}}},
\end{equation}
where $K_{\mathit{i}}$ is a constant in each segment and $\Gamma_{\mathit{i}}$ is the adiabatic index.

\begin{figure}
\centering
\includegraphics[width=1.0\columnwidth]{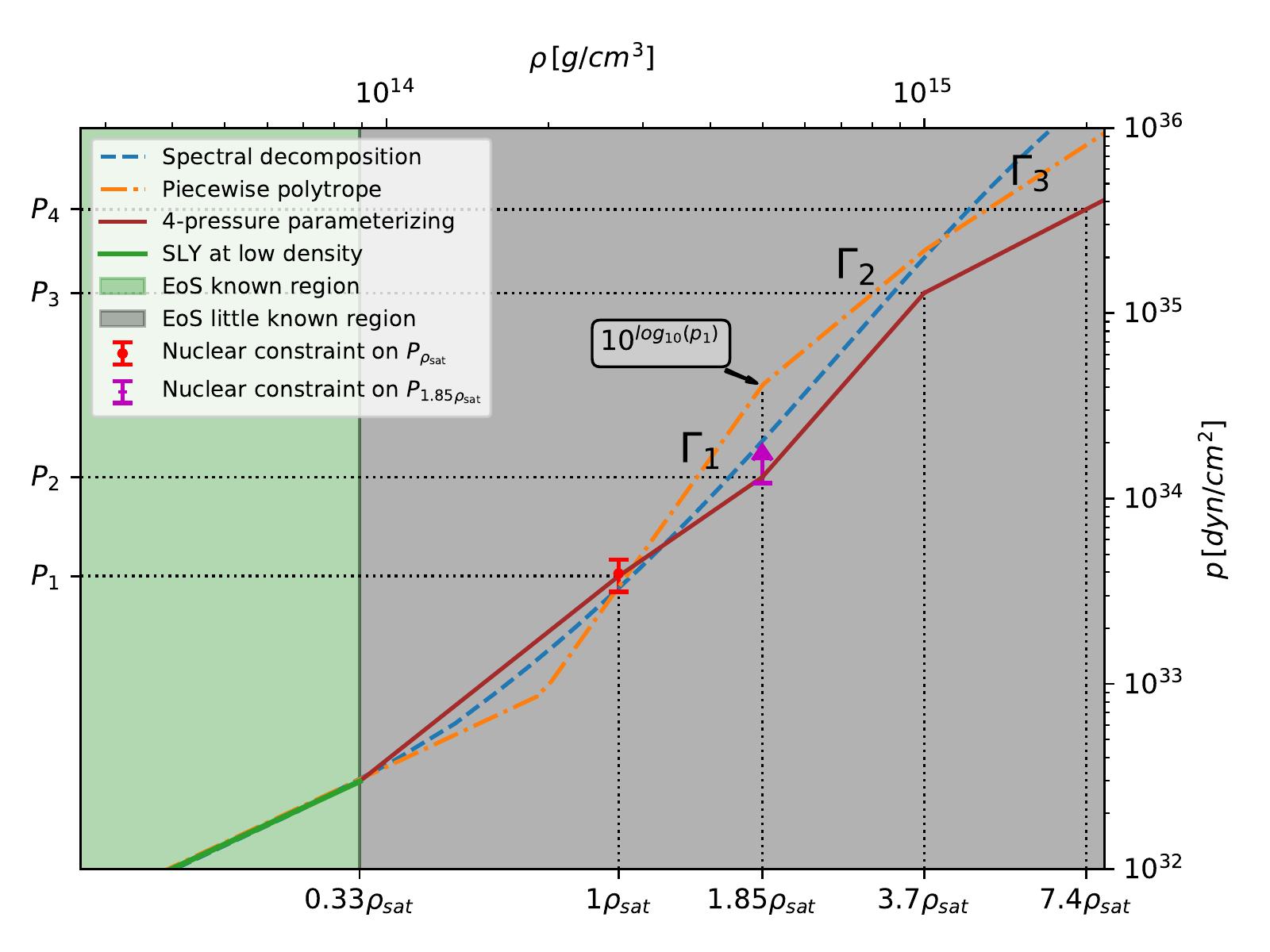}
\caption{Representative EoSs of different parameterizing methods in the form of pressure vs. rest-mass density. The green line represents the low density region of the EoS SLY \citep{2001A&A...380..151D}. The dashed blue line, the dashed-dotted orange line, and the solid brown line are obtained using the spectral decomposition method, the piecewise polytrope method, and the four-pressures parameterizing method, respectively. The uncertainties of $P_{\rho_{\rm sat}}$, $P_{1.85\rho_{\rm sat}}$ are obtained by nuclear theories/experiments \citep{2013ApJ...771...51L, 2017ApJ...848..105T}, which have been adopted by \citet{2016ApJ...820...28O} and \citet{Jiang2019}.}
\label{fig:para_methods}
\hfill
\end{figure}

The third method, which is based on the work of \citet{2009PhRvD..80j3003O} and \citet{2014PhRvD..89f4003L}, adopts four pressures $\{P_1, P_2, P_3, P_4\}$ at the corresponding densities of $\{1, 1.85, 3.7, 7.4\}\rho_{\rm sat}$ to parameterize the EoS ($P_2$ is the same as $p_1$ denoted in the second method). Similar to the second method, for each piece within the two nearby density boundaries, $p(\rho)$ also takes the polytrope form. The representative EoSs constructed from each method are shown in Fig.\ref{fig:para_methods}.

With a specific parameterizing method in hand, and assuming the two components of BNS share the same EoS, we can use the EoS parameters $\vec{\theta}_{\rm EOS}$ (e.g., $\{\gamma_0, \gamma_1, \gamma_2, \gamma_3\}$) to map the source frame mass $m_{\rm i}^{\rm src}(i=1,2)$ to the tidal deformability $\Lambda_{\rm i}(m_{\rm i}^{\rm src}, \vec{\theta}_{\rm EOS})$ of each star via solving the TOV equations and the Regge\textendash Wheeler equation. The source frame masses can be obtained from
\begin{equation}
\label{eq:mcqtom12}
\begin{split}
m_1^{\rm src}=q^{2/5}(1+q)^{1/5}\mathcal{M}_{\rm c}/(1+z_{\rm c}), \\
m_2^{\rm src}=q^{-3/5}(1+q)^{1/5}\mathcal{M}_{\rm c}/(1+z_{\rm c}),
\end{split}
\end{equation}
where $z_{\rm c}$, $q$ and $\mathcal{M}_{\rm c}$ are cosmological redshift, mass ratio and chirp mass of the BNS system, respectively.

\subsection{\label{sec:Bayesian}Bayesian Inference}
To constrain the parameter space of the EoS with GW data, we need to estimate the posterior probability density function (PDF) $p(\vec{\theta}_{\rm GW}|d)$, given the LIGO and Virgo data $d$, where $\vec{\theta}_{\rm GW}$ represents the detector frame parameters of the source. Through application of Bayes' Theorem, this posterior PDF is proportional to the product of the prior PDF $p(\vec{\theta}_{\rm GW})$ and the likelihood $L(d|\vec{\theta}_{\rm GW})$ of observing data $d$ given the waveform model described by $\vec{\theta}_{\rm GW}$,
\begin{equation}
\label{eq:Bayes}
p(\vec{\theta}_{\rm GW}|d) \propto L(d|\vec{\theta}_{\rm GW})p(\vec{\theta}_{\rm GW}).
\end{equation}
If we assume stationary Gaussian noise, then the log-likelihood of the single detector usually takes the function form,
\begin{equation}
\label{eq:Likelihood}
{\log L}(d|\vec{\theta}_{\rm GW}) = -2 \int_{0}^{\infty} \frac{|d(f) - h(\vec{\theta}_{\rm GW},f)|^2}{S_{\rm n}(f)} df + C,
\end{equation}
$S_{\rm n}(f)$, $d(f)$, and $h(\vec{\theta}_{\rm GW},f)$ are the one-sided power spectral density (PSD) of the noise, the Fourier transform of the time domain signal, and the frequency domain waveform generated using parameter $\vec{\theta}_{\rm GW}$, respectively.

For the case of GW170817, we fix the source location to the known position (${\rm R.A.}=197.450374^{\circ}$, ${\rm Decl.}=-23.381495^{\circ}$, $z_{\rm c}=0.0099$) as determined by electromagnetic (EM) observations \citep{2017ApJ...848L..12A, 2017ApJ...848L..28L} following \citet{2019PhRvX...9a1001A}, and assume the spin of each NS is aligned with the orbital angular momentum. We further marginalize the likelihood over coalescence phase (assuming that the signal is dominated by the $l = 2$, $|m| = 2$ modes) and the luminosity distance to accelerate Nest sampling \citep{2012PhRvD..85l2006A, 2019PhRvX...9a1001A, 2019EPJA...55...50R, 2019PASA...36...10T}. Thus the parameters of GW can take the form $\vec{\theta}_{\rm GW} = \{\Lambda_1(m_1^{\rm src},\vec{\theta}_{\rm EOS}),\Lambda_2(m_2^{\rm src},\vec{\theta}_{\rm EOS})\} \cup \{\mathcal{M}_{\rm c}, q, \chi_{\rm 1z}, \chi_{\rm 2z}, \theta_{\rm jn}, t_{\rm c}, \Psi\}$, where $\Lambda_1$($\Lambda_2$), $m_1^{\rm src}$($m_2^{\rm src}$), $ \chi_{\rm 1z}$($\chi_{\rm 2z}$), $\theta_{\rm jn}$, $t_{\rm c}$, and $\Psi$ are dimensionless tidal deformability, component masses in source frame, aligned spins, inclination angle, geocentric GPS time of the merger, and polarization of GW, respectively.

We take into account the cleaned 4096 Hz GW data \citep{2015JPhCS.610a2021V} whose GPS time spanning (1187008682, 1187008890)s, to avoid any contamination produced by the tapering effect after GPS time 1187008900 in the LIGO Hanford detector \citep{2018PhRvL.121i1102D}. We also adopt the publicly available PSD\footnote{\url{https://doi.org/10.7935/KSX7-QQ51}} \citep{2019PhRvX...9c1040A} and waveform model {\sc IMRPhenomD\_NRTidal} \citep{2017PhRvD..96l1501D}.

As for the priors of $\vec{\theta}_{\rm GW}$, a uniform distribution of $\cos{\theta_{\rm jn}}$ is adopted, and $\mathcal{M}_{\rm c}, q, \chi_{\rm 1z}, \chi_{\rm 2z}, t_{\rm c}, \Psi$ distribute uniformly in the range $[1.184,2.186]M_{\odot}$, $[0.5,1.0]$, $[-0.05,0.05]$, $[-0.05,0.05]$, $[1187008882, 1187008883){\rm s}$, $[0,2\pi] $, respectively.
\begin{table}[]
\begin{ruledtabular}
\centering
\caption{Priors of $\vec{\theta}_{\rm EOS}$ for Different Parameterizing Methods}
\label{tb:priors}
\begin{tabular}{ccc}
Methods                                                      & Parameters                                                    & Distributions \\ \hline
\multirow{4}{*}{Spectral}                              & $\gamma_0$                                                 & U\tablenotemark{\tiny{a}}(0.2, 2.0) \\
                                                                    & $\gamma_1$                                                 & U(-1.6, 1.7)            \\
                                                                    & $\gamma_2$                                                 & U(-0.6, 0.6)            \\
                                                                    & $\gamma_3$                                                 & U(-0.02, 0.02)       \\ \hline
\multirow{4}{*}{Piecewise}                           & $\log_{10}(p_1/{\rm dyn\,cm^{-2}})$               & U(34.1, 35.4)        \\
                                                                    & $\Gamma_1$                                                & U(0.6, 4.5)            \\
                                                                    & $\Gamma_2$                                                & U(0.6, 4.5)            \\
                                                                    & $\Gamma_3$                                                & U(0.6, 4.5)            \\ \hline
\multirow{4}{*}{Four-pressure}                    & $P_1/(10^{33}{\rm dyn\,cm^{-2}})$               & U(3.12, 4.7)           \\
                                                                    & $P_2/(10^{34}{\rm dyn\,cm^{-2}})$               & U(1.21, 8.0)           \\
                                                                    & $P_3/(10^{35}{\rm dyn\,cm^{-2}})$               & U(0.6, 7.0)             \\
                                                                    & $P_4/(10^{36}{\rm dyn\,cm^{-2}})$               & U(0.3, 4.0)             \\
\end{tabular}
\tablenotetext{\tiny{a}}{ U means uniform distribution.}
\end{ruledtabular}
\end{table}
And the EoSs parameterized by all the methods satisfy the following conditions: 1) causality constraint; 2) thermal stability $d\epsilon/dp > 0$; 3) nuclear constraint $P_{\rho_{\rm sat}} \in [3.12, 4.70]\times10^{33}~{\rm dyn\,cm^{-2}}$ and $P_{1.85\rho_{\rm sat}} > 1.21\times10^{34}~{\rm dyn\,cm^{-2}}$ \citep{2013ApJ...771...51L, 2016ApJ...820...28O, 2017ApJ...848..105T, Jiang2019}. Additionally, we limit the adiabatic index $\Gamma(p)$ to lie in the range of [0.6, 4.5] for spectral decomposition. Considering the above constraints, we choose reasonable ranges for the priors of $\vec{\theta}_{\rm EOS}$ of the given parameterizing method, e.g., the priors adopted by \citet{2018PhRvL.121p1101A} for the spectral method, the priors of \citet{Jiang2019} in their ${\rm Test E}$ for the four-pressure method, and compatible priors adopted by \citet{2018PhRvD..98f3004C} for the piecewise method. Thus the choice of our priors (summarized in Table.\ref{tb:priors}) can encompass a wide variety of candidate EoSs.

\subsection{\label{sec:mass-infer}Estimating the Masses of INSs with $z_{\rm g}$}
The maximum gravitational mass of nonrotating NSs ($M_{\rm TOV}$) can set a stringent constraint on the EoS parameter space \citep{2009PhRvD..79l4032R}. So far, though the actual value of $M_{\rm TOV}$ still remains uncertain, its range has been tightly bounded by the measurement of the massive neutron star (MNS) with mass $2.14^{+0.10}_{-0.09}M_{\odot}$ \citep[\objectname{PSR J0740+6620}]{2019NatAs.tmp..439C} and by the theoretical studies. In the so-called supramassive NS model for the peculiar X-ray afterglow of some short GRBs, $M_{\rm TOV}\sim 2.3M_{\odot}$ was suggested \citep{2013PhRvD..88f7304F}. An upper bound of $M_{\rm TOV}$ was set to $\sim 2.17M_{\odot}$, if a black hole central engine for GRB 170817A was formed quickly, though not promptly, after the double NS merger of GW170817 \citep[]{2017ApJ...850L..19M, 2017PhRvD..96l3012S, 2018ApJ...858...74M, 2018ApJ...852L..25R}. Very recently, considering that the angular momentum of the merger remnant of GW170817 was partly ``carried away" via the neutrino emission, GW radiation, the mass ejection, and the accretion disk, the $M_{\rm TOV}$ was suggested to be within the range of $[2.06,~2.25]M_{\odot}$ \citep{2019PhRvD.100b3015S, Shao2019}. Therefore, following \citet{2019arXiv190810352C}, we adopt the range of $M_{\rm TOV} \in [2.04, ~2.30]M_{\odot}$ (the lower limit is slightly smaller than the $68.3\%$ lower limit of \objectname{PSR J0740+6620} due to its quick rotation) to construct the EoSs using different parameterizing methods.

\begin{figure}
\centering
\includegraphics[width=1.0\columnwidth]{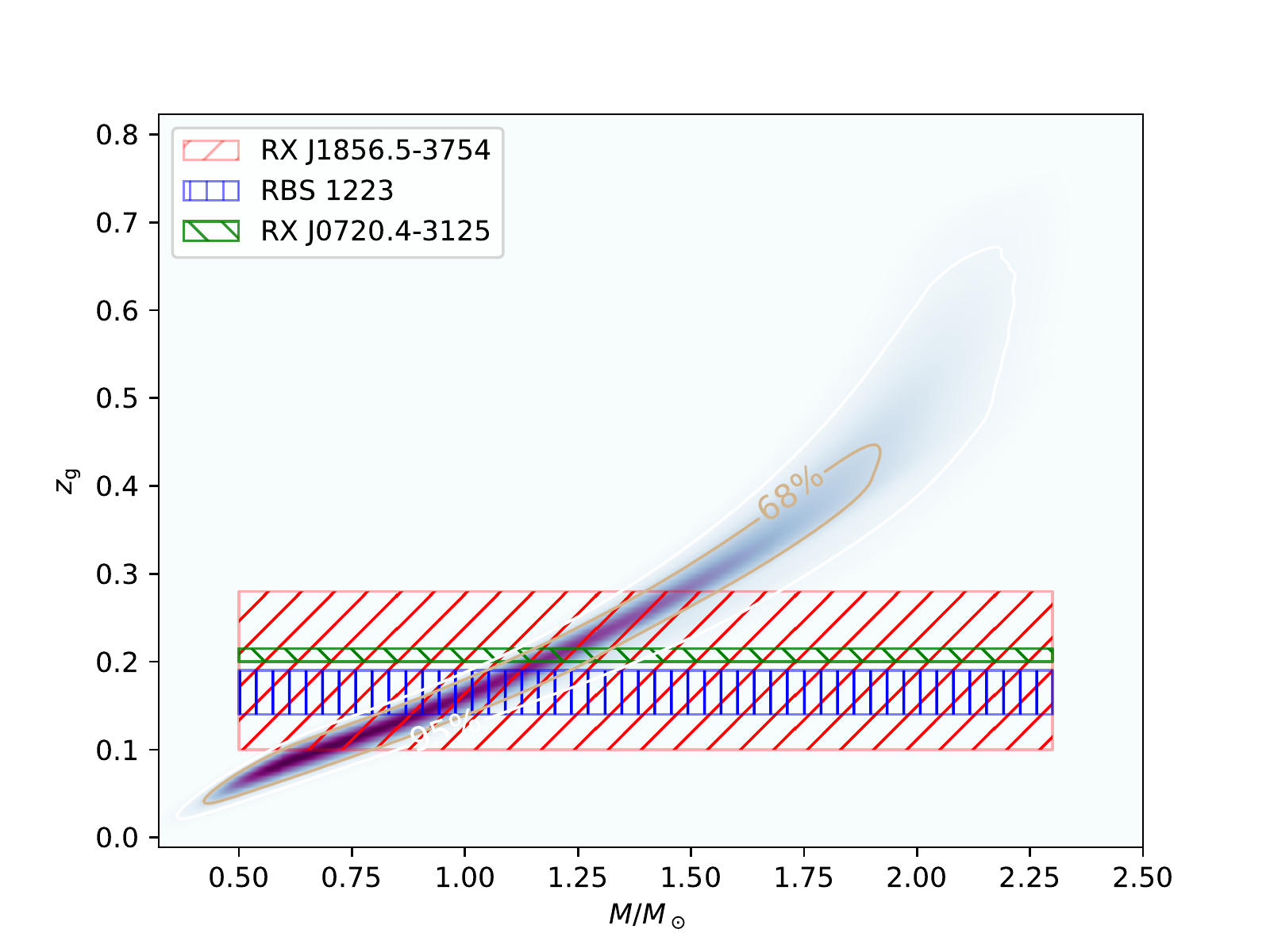}
\caption{Relationship between the masses and gravitational redshifts of NSs. The shadow areas represent the ranges of measured gravitational redshifts for the three sources, and the density plot is drawn using the piecewise EoS set, with masses of each EoS uniformly spaced.}
\label{fig:z_Mcurves}
\hfill
\end{figure}

To infer the masses of INSs using the EoS sets predicted by GW data, nuclear experiments, and the bounds on $M_{\rm TOV}$, we provide a novel approach described below. The measured gravitational redshift of an NS can be transformed to its compactness; therefore, the neutron-star mass can be obtained once the relationship of $M(R)$ is known. For each EoS, varying the central pressure or pseudo enthalpy, we can draw a curve in the $M\textendash R$ plane. Then, it is straightforward to have the $z_{\rm g}\textendash M$ curves using
\begin{equation}
z_{\rm g} = \frac{1}{\sqrt{1-\frac{2GM}{Rc^2}}}-1,
\end{equation}
where $G$ is the gravitational constant and $c$ is the speed of light. Then, 1000 values of $z_{\rm g}$ from the measured distribution of gravitational redshift (for instance, the accurate measurement of $z_{\rm g}=0.205^{+0.006}_{-0.003}$ for the source \objectname{RX J0720.4-3125} \citep{2017A&A...601A.108H}) are sampled, and for each we find the corresponding mass range using the $z_{\rm g}\textendash M$ curves of the whole EoS set as shown in Fig.\ref{fig:z_Mcurves}.

\section{\label{sec:results}Properties of NS and mass of INS}
\begin{figure}
\centering
\includegraphics[width=1.0\columnwidth]{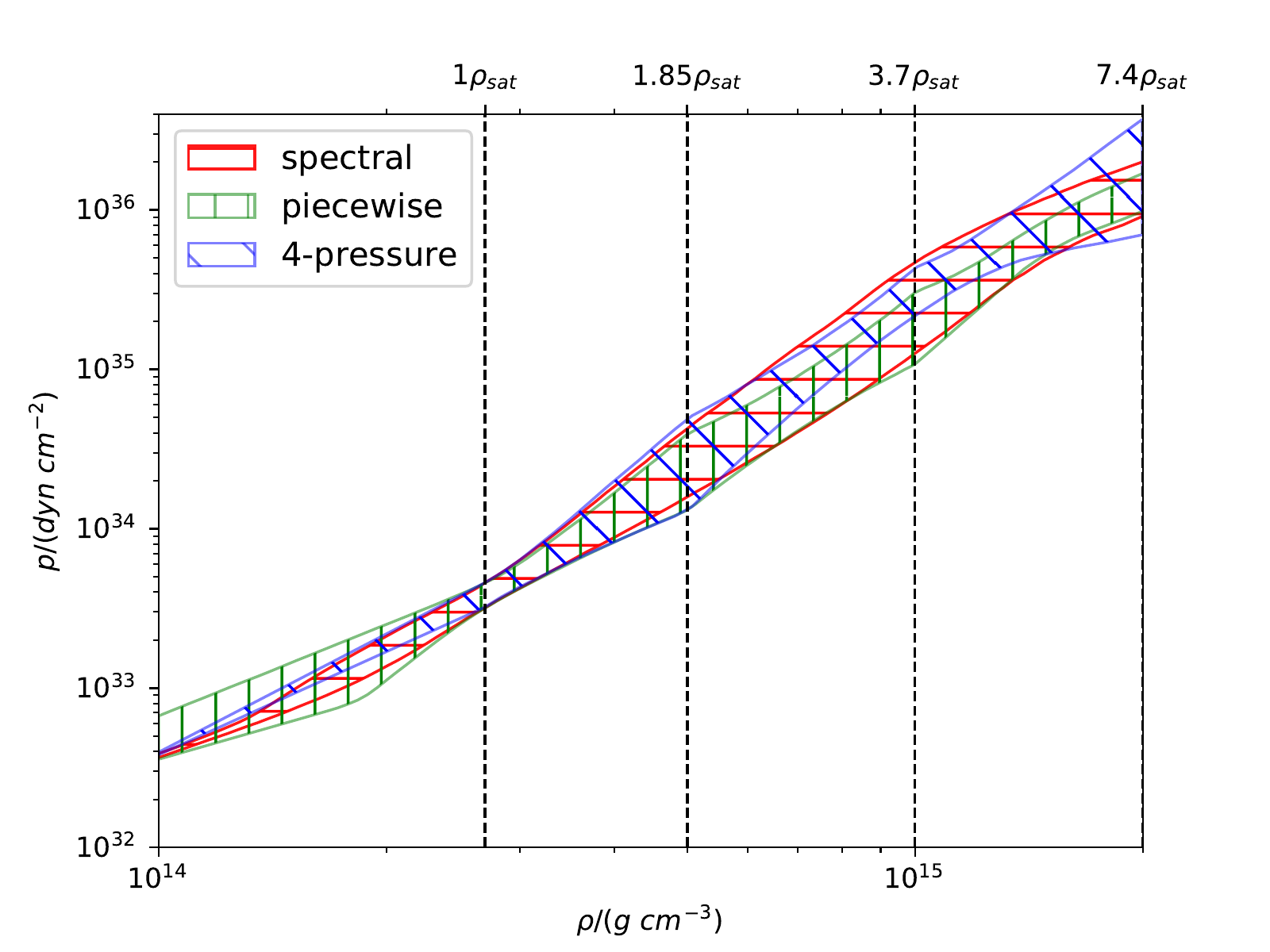}
\caption{The $90\%$ confidence regions of the EoS for three parameterizing methods.}
\label{fig:eos}
\hfill
\end{figure}

\begin{figure}
\centering
\includegraphics[width=1.0\columnwidth]{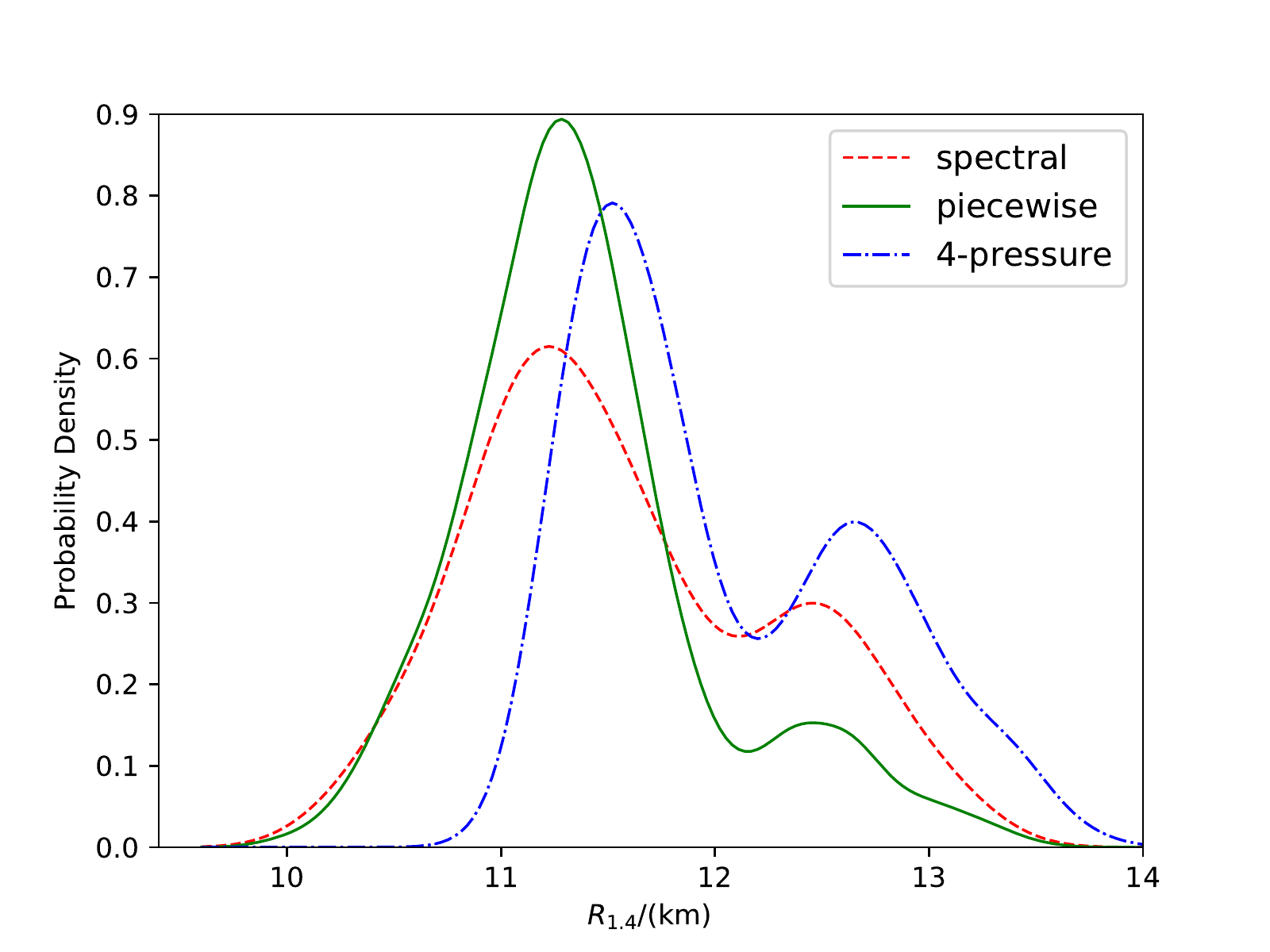}
\includegraphics[width=1.0\columnwidth]{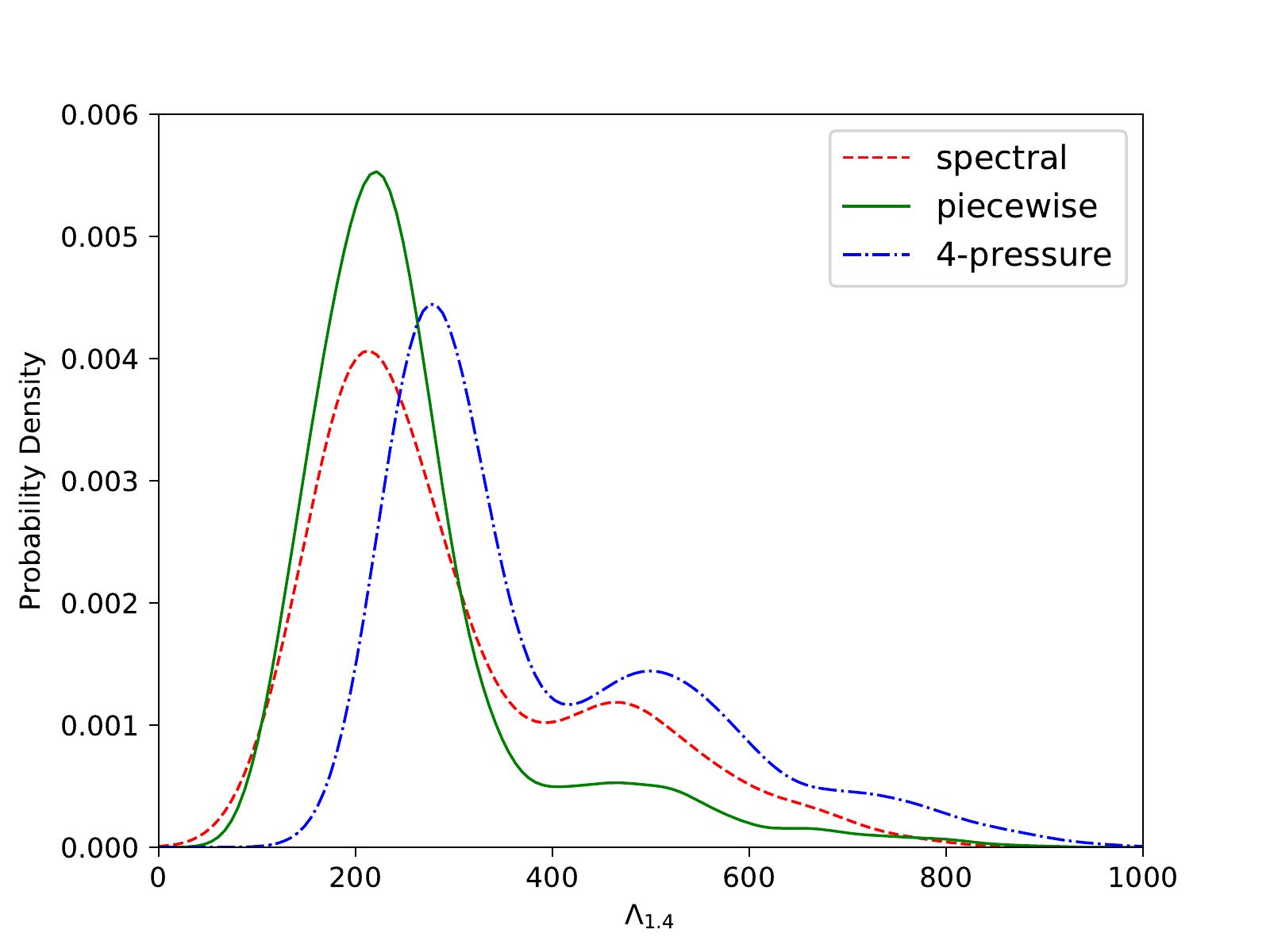}
\caption{Distributions of the bulk properties for the 1.4$M_{\odot}$ NS obtained from posterior samples of different parameterizing methods.}
\label{fig:macro}
\hfill
\end{figure}
The bulk properties, e.g., tidal deformability and radius of $1.4M_{\odot}$ NS, have been investigated by various groups \citep{2018PhRvL.120q2703A, 2018PhRvL.121i1102D, 2018PhRvL.120q2702F, 2018PhRvL.120z1103M, 2019PhRvX...9a1001A, 2019PhRvL.123n1101F, Jiang2019}. In this work, we reconstruct the EoS using GW data with nuclear constraints and the newly inferred $M_{\rm TOV}$ to obtain these properties. The $90\%$ confidence regions of EoS are shown in Fig.\ref{fig:eos}, and the bulk properties of NS are shown in Fig.\ref{fig:macro} and Table.\ref{tb:mtov_macro}.
\begin{table}
\centering
\begin{ruledtabular}
\caption{The $68\%$ Confidence Intervals of Bulk Properties for the 1.4$M_{\odot}$ NS Derived from Posterior Samples of Different Parameterizing Methods}
\begin{tabular*}{1.0\columnwidth}{cccc}
Properties/Methods                & Spectral     & Piecewise    & Four-pressure \\ \hline
$R_{\rm 1.4}/({\rm km})$ & $11.5_{-0.6}^{+1.0}$ & $11.3_{-0.4}^{+0.6}$ & $11.8_{-0.4}^{+1.0}$ \\
$\Lambda_{1.4}$    & $250_{-70}^{+220}$ & $230_{-60}^{+100}$ & $320_{-60}^{+240}$ \\
\end{tabular*}
\label{tb:mtov_macro}
\end{ruledtabular}
\end{table}
Our results show that the distributions of both $R_{1.4}$ and $\Lambda_{1.4}$ have bimodal behaviors, which are consistent with \citet{2018PhRvL.121p1101A} and \citet{Jiang2019}. As found in the data analysis of the two individual Advanced LIGO detectors by \citet{2018arXiv181206100N}, the probability distributions of tidal deformabilities ($\Lambda$) inferred from the data of H1 and L1 are different. The underlying physical reasons are that H1 has a higher detectability than L1 at frequencies above 100 Hz (therefore a better measurability of $\Lambda$ for H1; \citealt{2012PhRvD..85l3007D}) and the specific localization of GW170817 further increases the signal-to-noise ratio of H1 (the details will be presented in a dedicated investigation by \citealt{Han2019}).

Up to now, the masses of dozens of NSs in BNS or neutron stat-white dwarf (NSWD) systems have been measured (see the references in \citealt{2016ARA&A..54..401O}), and the mass distribution has been investigated in the literature \citep[e.g.,][]{2012ApJ...757...55O, 2013ApJ...778...66K, 2017ApJ...846..170T, 2018arXiv180403101H}. For INSs, their masses and mass distribution are unknown.
\begin{table}
\tiny
\centering
\begin{ruledtabular}
\caption{The $68\%$ Confidence Intervals of Masses for Three Sources Derived from the Measured $z_{\rm g}$}
\begin{tabular*}{1.0\textwidth}{ccccc}
Sources/Methods                  & Spectral     & Piecewise    & Four-pressure  & Average \\
\hline
\objectname{RX J1856.5-3754} & $1.20_{-0.30}^{+0.29}$ & $1.19_{-0.29}^{+0.28}$ & $1.28_{-0.28}^{+0.29}$ & $1.24_{-0.29}^{+0.29}$ \\
\objectname{RX J0720.4-3125} & $1.22_{-0.06}^{+0.10}$ & $1.21_{-0.04}^{+0.06}$ & $1.26_{-0.05}^{+0.11}$ & $1.23_{-0.05}^{+0.10}$ \\
\objectname{RBS 1223} & $1.08_{-0.11}^{+0.19}$ & $1.07_{-0.10}^{+0.19}$ & $1.11_{-0.12}^{+0.21}$ & $1.08_{-0.11}^{+0.20}$ \\
\end{tabular*}
\label{tb:masses}
\end{ruledtabular}
\end{table}
Fortunately, the X-ray spin phase-resolved spectroscopic study of some bright thermally emitting INSs and the fit of highly magnetized atmospheric models allow the community to estimate their compactness. Based on the multiepoch observations conduced by XMM-Newton, the gravitational redshifts of \objectname{RBS 1223}, \objectname{RX J0720.4-3125}, and \objectname{RX J1856.5-3754}, three members of the so-called ``The Magnificent Seven," have been determined to be $0.16^{+0.03}_{-0.02}$, $0.205^{+0.006}_{-0.003}$, and $0.22^{+0.06}_{-0.12}$ \citep{2014JPhCS.496a2015H, 2017A&A...601A.108H}. Applying the approach described in Sec.\ref{sec:mass-infer}, the masses of these three sources are evaluated. The $68\%$ confidence intervals are summarized in Table.\ref{tb:masses},
\begin{figure}
\centering
\includegraphics[width=1.0\columnwidth]{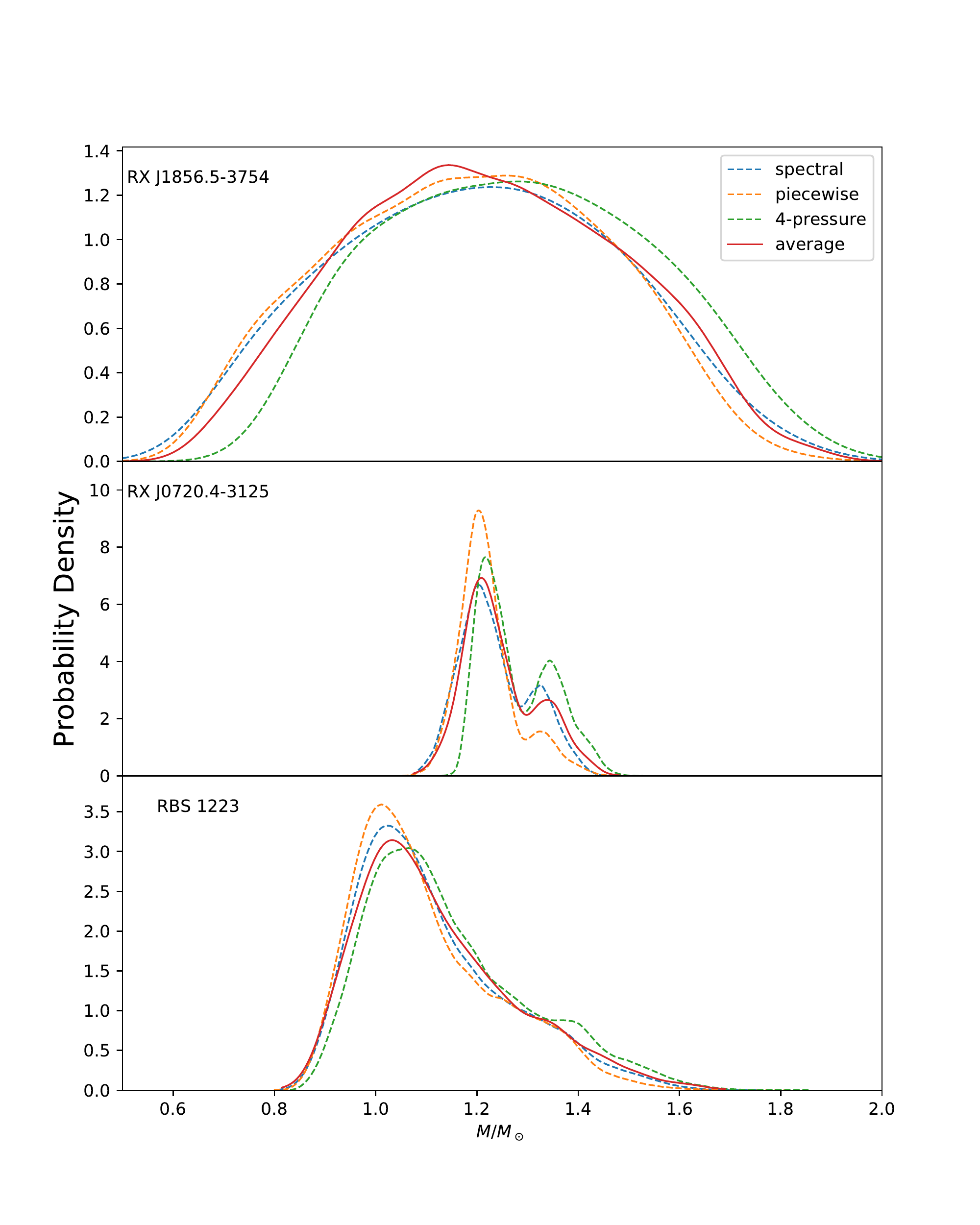}
\caption{Mass distributions of the three sources found in our approaches.}
\label{fig:masses}
\end{figure}
and the mass distributions of each source are shown in Fig.\ref{fig:masses}. The narrow mass distribution of \objectname{RX J0720.4-3125} is mainly governed by the uncertainty of the EoS constrained with the current data because of the very small error of the $z_{\rm g}$. While the wide mass distribution of \objectname{RX J1856.5-3754} reflects its large uncertainty of gravitational redshift measurement. Note that our results are consistent among different parameterizing methods, so we incorporate an equal number of samples from the posterior of each method to get our overall results for each source (a similar procedure was adopted by \citealt{2019PhRvX...9c1040A}).

\section{\label{sec:summary}Discussion and Summary}

Different from the NSs in the binary systems, the masses of the isolated NSs are hard to measure. In this work, which benefited from the significant progress made recently on constraining the EoS of NSs, we propose a novel approach to estimate the mass of an NS with a known gravitational redshift (see Sec.\ref{sec:mass-infer} and Fig.\ref{fig:z_Mcurves}). In our approach the influence on bulk properties of NSs by adopting different parameterized EoSs has been examined. We implement the generally used parameterizing methods, including the pressure-based spectral decomposition, piecewise polytrope, and four-pressure parameterization, to construct the EoS sets by analyzing GW data with additional constraints, including the pressure range limited by nuclear theories/experiments and maximum mass of nonrotating NSs. Then, we map the measured gravitational redshift to $z_{\rm g}\textendash M$ curves derived from the samples of the constructed EoS sets, and get the corresponding masses of INSs for three sources \objectname{RX J1856.5-3754}, \objectname{RX J0720.4-3125}, and \objectname{RBS 1223} in M7, which are $1.24_{-0.29}^{+0.29}M_{\odot}$, $1.23_{-0.05}^{+0.10}M_{\odot}$, and $1.08_{-0.11}^{+0.20}M_{\odot}$, respectively. The masses derived from different EoS sets with the three parameterizing methods are consistent with each other, which suggests that our result is reliable. We caution that these inferred masses cannot be combined with the gravitational redshifts to further constrain EoS, because they are not independent of each other. It is different from the measurements of the masses and gravitational redshifts or radii of NSs carried out by other channels, e.g., $M\textendash R$ obtained using LMXBs' data or $z_{\rm g}\textendash M$ observed with NICER, which are expected to tightly constrain the EoS \citep{2019ApJ...881...73W}. 

\begin{figure}
\centering
\includegraphics[width=0.8\columnwidth]{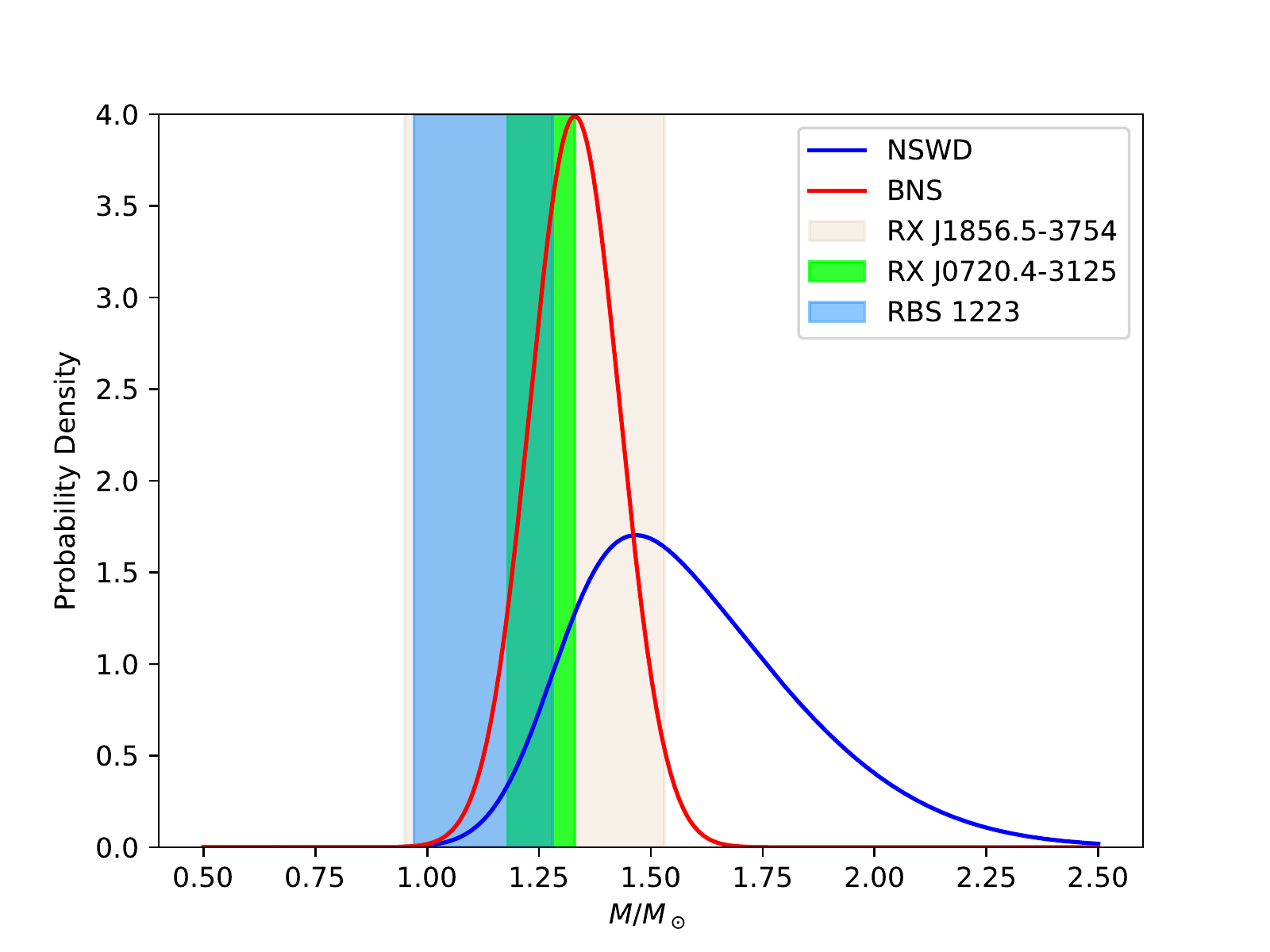}
\caption{Mass distributions of NSs in BNS and NSWD systems. The shadow area of each source represents $1\sigma$ uncertainty of the mass averaged among the three methods.}
\label{fig:massdistri}
\hfill
\end{figure}
We also compare our results with the mass distributions of NSs in BNS and NSWD systems \citep{2013ApJ...778...66K}, as shown in Fig.\ref{fig:massdistri}. The masses of INSs are more similar to that of BNS systems, which may indicate that there was no significant mass accretion onto NSs and the measured values trace initial masses when they were born. In the future, with the detection of BNS or NSBH systems by the LIGO/Virgo Collaboration and the observations of NICER, the parameter space of the EoS is expected to reduce to a narrow range \citep{2019PhRvD.100j3009H}. With our approach and the increasing samples of INSs with gravitational redshift measurements, the mass distribution of such objects can be reconstructed and their evolutionary paths will be better understood. For example, a constructed mass distribution can help to unify the apparent diversity of the classes of INSs \citep{2013MNRAS.434..123V, 2015SSRv..191..171P}, or help to determine the minimum mass forming an NS \citep{2018MNRAS.481.3305S}; and the initial mass function may be used to check the theoretical expectations for remnant masses produced by electron-capture versus Fe-core collapse SNe \citep{2004ApJ...612.1044P,2013ApJ...778...66K}.

\begin{figure}
\centering
\includegraphics[width=1.0\columnwidth]{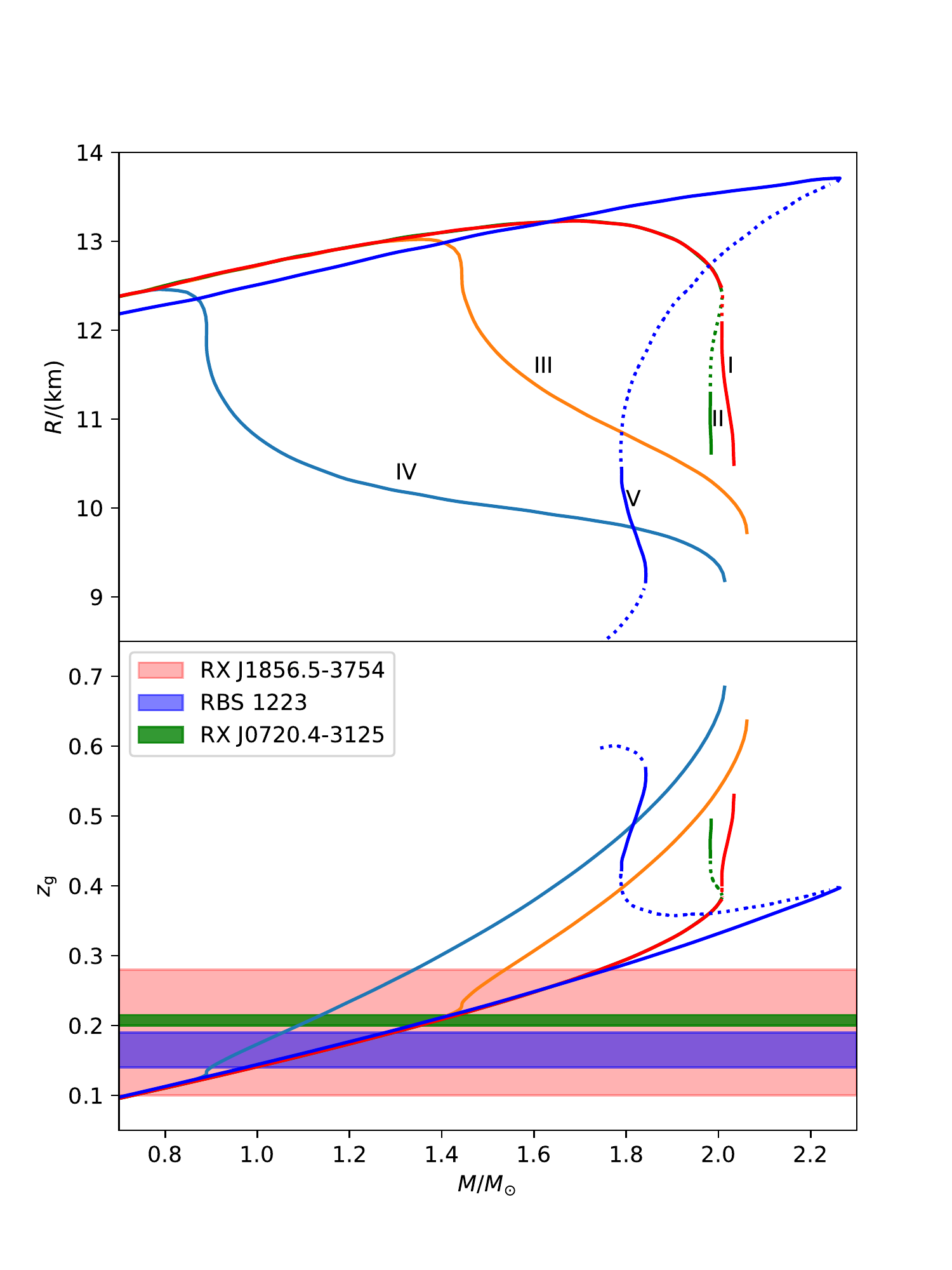}
\caption{The $M\textendash R$ and $M\textendash z_{\rm g}$ relations of a few representative phase-transition EoSs adopted from the literature \citep{2018PhRvL.120z1103M, 2019PhRvD..99j3009M}. The dashed lines are unstable branches, lines labeled with Roman numerals are twin-star branches, and shadow areas are the measured gravitational redshifts of the three sources.}
\label{fig:phase}
\hfill
\end{figure}
Please note that our investigations are based on the assumption of the absence of phase transition. There are many works examining the existence of a hybrid hadron-quark star whose EoS undergoes a phase transition \citep{2018PhRvL.120q2703A, 2018PhRvL.120z1103M, 2018PhRvC..98d5804T, 2019PrPNP.10903714B, 2019PhRvD..99j3009M}. Such phase transitions may soften or stiffen the EoS to different degrees, and thus would modify the $M\textendash R$ relationship. Among some configurations of the phase transition, there may exist the scenarios of twin-star \citep{2000A&A...353L...9G, 2002NuPhA.702..217F, 2002PhRvL..89q1101S} or two-family hybrid stars \citep{2018ApJ...852L..32D}. Shown in Fig.\ref{fig:phase} are some representative EoSs suffering from phase transitions. In the measured gravitational redshift ranges, if covered by the twin-star branches, we will get a smaller $M$. This is understandable because $z_{\rm g}$ fixes the compactness, and a small radius points toward a small mass. However, if covered by the normal branches, the masses will be slightly enhanced since the phase transition will allow stiffer EoSs to satisfy the maximum mass constraints. Thus the masses inferred with $z_{\rm g}$ will be influenced by the ratio of twin-star branches to the normal branches, which depends on the parameter configurations of $\Delta e\textendash p_{\rm tr}$ that describe the discontinuity of EoS. Using the data of GW170817 and constraints of $M_{\rm TOV}$, \citet{2018PhRvL.120z1103M} showed that this ratio is $\sim 2\%$. The rapidly increasing sample of double NS merger events will considerably tighten the constraints on the possible phase transition. Then a more robust evaluation of $M$ of the NS(s) with a well measured $z_{\rm g}$, as outlined in this work, will be achieved.

\acknowledgments
We thank the anonymous referee for the helpful suggestions. This work was supported in part by NSFC under grants of No. 11525313 (i.e., Funds for Distinguished Young Scholars), No. 11921003, No. 11433009, and No. U1738126, the Chinese Academy of Sciences via the Strategic Priority Research Program (grant No. XDB23040000), and the Key Research Program of Frontier Sciences (No. QYZDJ-SSW-SYS024). This research has made use of data and software obtained from the Gravitational Wave Open Science Center (\url{https://www.gw-openscience.org}), a service of LIGO Laboratory, the LIGO Scientific Collaboration and the Virgo Collaboration. LIGO is funded by the U.S. National Science Foundation. Virgo is funded by the French Centre National de Recherche Scientifique (CNRS), the Italian Istituto Nazionale della Fisica Nucleare (INFN) and the Dutch Nikhef, with contributions by Polish and Hungarian institutes.

\software{PyCBC \citep[version 1.13.6, ascl:1805.030, \url{https://github.com/gwastro/pycbc}]{2018ascl.soft05030T}, Bilby \citep[version 0.5.5, ascl:1901.011, \url{https://git.ligo.org/lscsoft/bilby/}]{2019ascl.soft01011A}, PyMultiNest \citep[version 2.6, ascl:1606.005, \url{https://github.com/JohannesBuchner/PyMultiNest}]{2016ascl.soft06005B}, LALSuite \citep[version 6.57, \url{https://doi.org/10.7935/GT1W-FZ16}]{lalsuite}}

\end{document}